\documentclass{article}
\usepackage{amssymb}
\usepackage{amsmath,amsthm}

\newtheorem{theorem}{Theorem}[section]

\newtheorem{proposition}[theorem]{Proposition}

\newtheorem{lemma}[theorem]{Lemma}
\newtheorem{remark}[theorem]{Remark}

\renewenvironment{enumerate}{\begin{list}{{\bf(\roman{enumi})}}{\usecounter{enumi}}}{\end{list}}
\renewcommand{\leq}{\leqslant}
\renewcommand{\geq}{\geqslant}
\def\N{{\mathbb N}}

\def\Nz{{\mathbb N}_0}

\def\Fq{{\mathbb F}_q}

\newcommand\weight[1]{{\bf #1}}
\def\phi{\varphi}
\def\Goto{\longmapsto}

\title{Redundancies of Correction-Capability-Optimized Reed-Muller Codes}
\date{\today}
\author{Maria Bras-Amor\'os\thanks{Universitat Aut\`onoma de
Barcelona, mbras@deic.uab.cat}\ \ and Michael E. O'Sullivan\thanks{San
Diego State University, mosulliv@sciences.sdsu.edu}}

\begin{document}
\maketitle

\begin{abstract}
This article is focused on
some variations of Reed-Muller codes
that yield improvements to the rate for a prescribed  decoding
performance under the Berlekamp-Massey-Sakata algorithm with majority
voting.  Explicit formulas for  the redundancies
of the new codes are given.
\end{abstract}

%\begin{keywords}
%Reed-Muller code, Feng-Rao improved code.
%\end{keywords}

\section*{Introduction}

Reed-Muller codes belong to the family of evaluation codes, commonly
defined on an order domain. The decoding algorithm widely used for
evaluation codes is an adaptation of the Berlekamp-Massey-Sakata
algorithm together with the majority voting algorithm of
Feng-Rao-Duursma.
By analyzing majority voting,
one realizes that only some of the parity checks
are really necessary to perform correction of a given number of errors.
New codes can be defined with just these few checks, yielding larger dimensions
while keeping the same correction capability as standard codes \cite{FeRa:improved,HoLiPe:agc}.
These codes are often called Feng-Rao improved codes.

A different improvement to standard evaluation codes is given in
\cite{O'Sullivan:hermite-beyond}.
The idea is that under the
Berlekamp-Massey-Sakata algorithm with majority voting, error vectors
whose weight is larger than half the minimum distance of the code are
often correctable.  In particular this occurs for {\it generic errors}
(also called independent errors in
\cite{Pellikaan:independent_errors,JeNiHo}), whose  technical
algebraic definition  can be found in the mentioned references.
Generic  errors of weight $t$ can be a very large proportion of all
possible errors of weight $t$, as in  the case of the examples worked out in
\cite{O'Sullivan:hermite-beyond}.
This suggests that a code be designed to correct only {\it
generic errors}  of weight $t$ rather than  all error words of
 weight $t$.
Using this
restriction, one obtains new codes with much
larger dimension than that of standard evaluation codes correcting
the same number of errors.

In \cite{BrOS:AAECC} both ideas are combined. Minimal order subsets
are accurately designed in order to ensure correction capability of
$t$ generic errors, under the Berlekamp-Massey-Sakata algorithm with
majority voting.

The scope of this work is to give explicit formulae for the redundancies
of all the Reed-Muller improved codes.
In Section~\ref{sec:defs}
we recall the definitions of the correction-capability-optimized
codes. In Section~\ref{sec:red}
we give formulas to find their redundancies.

\section{Correction-capability-optimized Reed-Muller codes}
\label{sec:defs}

Let $n=q^m$ and call $P_1,\ldots,P_n$ the $n$ points in $\Fq^m$.
Let
$\Fq[x_1,\ldots,x_m]_{\leq s}$ be the subspace of
$\Fq[x_1,\ldots,x_m]$ of polynomials with total degree $\leq s$
and let $\varphi_s$ be the map $\Fq[x_1,\ldots,x_m]_{\leq s} \longrightarrow
\Fq^n$, $f \Goto {(f(P_1),\ldots,f(P_n))}.$
The
{\it Reed-Muller code} {\bf $RM_q(s,m)$} is defined as
the orthogonal space of
the image of~$\varphi$.

Let $A=\Fq[x_1,\dots,x_m]$ and let $\varphi:f \mapsto
 (f(P_1),\ldots,f(P_n))$.
Variations of Reed-Muller codes can be defined by means of
a subset $W$ of monomials in $\Fq[x_1,\dots,x_m]$.
The order-prescribed Reed-Muller code associated to $W$
is $$C_W=<\varphi(W)>^\perp.$$

Let $\ll$ be the graded lexicographic order
on monomials in $A$ with $x_m\ll x_{m-1}\ll \dotsb \ll x_1$.
Let $z_i $ be the $i$-th monomial with respect to $\ll$, starting
with $z_0=1$.
Let $j$ be such
that $z_j=x_1^s$, then $z_{j+1}=x_m^{s+1}$ and
$\Fq[x_1,\ldots,x_m]_{\leq s}$ is the space generated by $\{z_i: i\leq
j\}$. Consequently we have $$RM_q(s,m)=C_{\{z_i: i\leq j\}}.$$
More generally, one can define the  {\it standard Reed-Muller code}
for any given $j $ to be $C_{\{z_i: i\leq j\}}$.

For $m\in\Nz$ let
$$\nu_m=\lvert \{j\in \N_0: z_j\mbox{ divides } z_m\}\rvert.$$
The sequence given by the values $\nu_i$ with $i\in\N_0$ has two
important applications.
On the one hand, it is used to define
bounds on the minimum distance of evaluation codes
\cite{FeRa:dFR,KiPe:telescopic,HoLiPe:agc}.
On the other hand it is used to design Feng-Rao improved codes \cite{FeRa:improved,HoLiPe:agc}.
The main results used for defining correction-capability-optimized
codes are the two following lemmas.

\begin{lemma}
\cite{FeRa:improved}
All errors of weight $t$ can be corrected by $C_W$  if $W$
contains all monomials $z_i$ with $\nu_i<2t+1$.
\end{lemma}

\begin{lemma}
\cite{BrOS:AAECC}
All generic errors of weight $t$ can be corrected by $C_W$  if $W$
contains all monomials $z_i$ which are not a product $z_jz_k$ for any $j,k\geq t$.
\end{lemma}

\paragraph{Standard Reed-Muller codes} To design a standard Reed-Muller
code
which will correct
$t$ errors, let $m(t)  = \max\{i\in\Nz : \nu_i< 2t+1\}$.
Let $R(t) = \{ z_i: i\leq m(t) \}$ and $r(t)=\lvert R(t)\rvert$.  The code
$C_{R(t)}$ has minimum distance at least $2t+1$.

\paragraph{Feng-Rao improved codes}  To design an order-prescribed
Reed-Muller code correcting $t$ errors, we take
$\widetilde{R}(t)= \{z_i : \nu_i< 2t+1\}$ and use the code
$C_{\widetilde{R}(t)}$.
Let $\widetilde{r}(t)=\lvert \widetilde{R}(t) \rvert=m(t)+1$.
The Feng-Rao improved Reed-Muller code correcting $t$ errors requires
$r(t)-\widetilde{r}(t)$ fewer check symbols than the standard Reed-Muller code
correcting $t$ errors.

\paragraph{Standard generic Reed-Muller codes} To design a standard Reed-Muller
code
that will correct
all generic errors of weight at most $t$, let
$m^*(t) = \max\{i: z_i\neq z_jz_k \mbox{ for all }j,k\geq t\}$.
Define $R^*(t) = \{ z_i : i \leq m^*(t)\}$.
The number of check symbols for the code $C_{R^*(t)}$ is
$r^*(t)=\lvert R^*(t)\rvert=m^*(t)+1$.

\paragraph{Improved generic Reed-Muller codes}  To design an order-prescribed
Reed-Muller code correcting $t$ generic errors, we use the code
$C_{\widetilde{R}^*(t)}$ where
$\widetilde{R}^*(t)$ is
$\{z_i: z_i\neq z_jz_k \mbox{ for all }j,k\geq t\}$.
Let $\widetilde{r}^*(t)=\lvert \widetilde{R}^*(t)\rvert$.  Clearly
$\widetilde{r}^*(t) \leq r^*(t)$.

\section{Explicit formulae for the redundancies}
\label{sec:red}

\begin{lemma}
\label{lemma: nui per rm} Suppose $z_i=x_1^{a_1}\cdot\dots\cdot
x_m^{a_m}$. Then, $\nu_i=\prod_{l=1}^m(a_l+1).$
\end{lemma}

\begin{proof}
It is obvious, since the monomial $x_1^{b_1}\cdot\dots\cdot x_m^{b_m}$
divides $x_1^{a_1}\cdot\dots\cdot x_m^{a_m}$ if and only if $0\leq b_l\leq
a_l$ for all $1\leq l\leq m$.
\end{proof}

The next proposition quantifies the redundancy of non-generic codes.
\begin{proposition}
\label{proposition:red-rm-1} For every $t\in\N_0$,
\begin{enumerate}
\item$r(t)=\binom{2t-1+m}{m} ,$
\item$\tilde{r}(t)=\lvert\left\{a\in\N_0^m:
\prod_{l=1}^m(a_l+1)< 2t+1\right\}\rvert.$
\end{enumerate}
\end{proposition}

\begin{proof}
\mbox{}
\begin{enumerate}
\item The monomial $z_s$ of largest lexicographic order
for which $\nu_s<2t+1$ is $z_s=x_1^{2t-1}$. Thus,
$m(t)=s=\binom{2t-1+m}{m}-1$ and $r(t)=\binom{2t-1+m}{m}$.
\item
It is a direct consequence of Lemma~\ref{lemma: nui per rm}.
\end{enumerate}
\end{proof}

We now give the redundancies for generic codes.

\begin{proposition}
\label{proposition:red-rm-2} Suppose $z_t=x_1^{a_1}\cdot\dots\cdot
x_m^{a_m}$ and let $\weight{a} = |a|= a_1+a_2+\dots+a_m$.
\begin{enumerate}
\item
If $a_1=a_2=\dots=a_{m-1}=0$ (hence $a_m= \weight{a}$), then
\begin{itemize}
\item $r^*(t)=
\binom{2\weight{a}-1+m}{m}$,
\item
$\tilde{r}^*(t)=r^*(t)$.
\end{itemize}
\item
Otherwise,
\begin{itemize}
\item
$r^*(t)=\binom{2\weight{a}+1+m}{m}- \sum_{k=1}^m\binom{2\weight{a}-
\sum_{l=1}^k a_l+m-k}{m-k},$
\item
$\tilde{r}^*(t)=r^*(t) -1-\sum_{k=1}^{m-1} \sum_{j=k}^{m-1}
\binom{2\weight{a}-2-\sum_{i=1}^j a_i -\sum_{l=1}^k a_l+m-j}{m-j}\\
-\lvert\{k : \weight{a}-\sum_{i=1}^k a_i>0\}\rvert.$
\end{itemize}
\end{enumerate}
\end{proposition}

\begin{proof}
\mbox{}
\begin{enumerate}
\item If $z_t=x_m^{a_m}$
then $\{z_i: z_i=z_jz_k, j,k\geq t\}=\{z_i: \deg z_i \geq 2a_m\}.$ So,
$r^*(t)=\tilde{r}^*(t)=\binom{2a_m-1+m}{m}.$

\item Otherwise,
$\{z_j: j\geq t\}= \{z_j: \deg(z_j)>\weight{a}\}\sqcup\{z_j:
\deg(z_j)=\weight{a}\mbox{ and }j\geq t\}.$ So,
\begin{eqnarray*}
\{z_jz_k: j,k\geq t\}&=&\{z_jz_k: \deg(z_jz_k)>2\weight{a}+1
\}\\
& &\sqcup \{z_jz_k: \deg(z_jz_k)=2\weight{a}+1\mbox{ and }
j,k\geq t\}\\
& &\sqcup \{z_jz_k:\deg(z_jz_k) =2\weight{a}\mbox{ and } j,k\geq t
\}.
\end{eqnarray*}

Let us introduce the following notation:
\begin{eqnarray*}
P_{2\weight{a}}&=&\{z_jz_k: \deg(z_jz_k)=2\weight{a} \mbox{ and }
j,k\geq t
\},\\
P_{2\weight{a}+1}&= &\{z_jz_k: \deg(z_jz_k)=2\weight{a}+1 \mbox{ and
} j,k\geq t \}.
\end{eqnarray*}

Then $\tilde{r}^*(t)=\lvert\{z_i:\deg(z_i)\leq
2\weight{a}+1\}\rvert-\lvert P_{2\weight{a}}\rvert-\lvert
P_{2\weight{a}+1}\rvert.$

One may verify  that  the monomial $z_i=x_1^{b_1}\cdot\dots\cdot
x_m^{b_m}$ with $\deg(z_i)=2\weight{a}$ is in $P_{2\weight{a}}$ if
and only if it satisfies one of the following:
\begin{itemize}
\item $b_l=2a_l$ for all $1\leq l\leq m$,
\item There exists $1\leq k\leq m-1$ such that
\begin{itemize}
\item
$b_l=2a_l$ for all $1\leq l\leq k-1$,
\item
$b_k\geq 2a_k+2$
\end{itemize}
\item There exists $1\leq k<j\leq m-1$ such that
\begin{itemize}
\item
$b_l=2a_l$ for all $1\leq l\leq k-1$,
\item
$b_k=2a_k+1$,
\item
$b_l=a_l$ for all $k+1\leq l\leq j-1$,
\item
$b_j\geq a_j+1$
\end{itemize}
\item There exists $1\leq k\leq m-1$ such that
\begin{itemize}
\item
$b_l=2a_l$ for all $1\leq l\leq k-1$,
\item
$b_k=2a_k+1$,
\item
$b_l=a_l$ for all $k+1\leq l\leq m-1$,
\item
$b_m\geq a_m$
\end{itemize}
\end{itemize}

Consequently,
\begin{eqnarray*}
\lvert P_{2\weight{a}}\rvert&=& 1+\sum_{k=1}^{m-1}
\binom{2\weight{a}-2-2\sum_{l=1}^k a_l+m-k}{m-k}
\\
& &+ \sum_{k=1}^{m-1} \sum_{j=k+1}^{m-1} \binom{2\weight{a}-2-
\sum_{l=1}^j a_l -\sum_{l=1}^k a_l +m-j}{m-j}
\\
& &+
\lvert\{k:\weight{a}-\sum_{l=1}^k a_l>0\}\rvert\\
&=& 1+\sum_{k=1}^{m-1} \sum_{j=k}^{m-1}
\binom{2\weight{a}-2-\sum_{l=1}^j a_l -\sum_{l=1}^k a_l+m-j}{m-j}
\\
& &+ \lvert\{k:\weight{a}-\sum_{l=1}^ka_l>0\}\rvert.
\end{eqnarray*}

Similarly, the monomial $z_i=x_1^{b_1}\cdot\dots\cdot x_m^{b_m}$
with $\deg(z_i)=2\weight{a}+1$ is in \break $P_{2\weight{a}+1}$ if
and only if there exists $k$, $1\leq k\leq m$, such that
\begin{itemize}
\item $b_l=a_l$ for all $1\leq l\leq k-1$,
\item $b_k\geq a_k+1.$
\end{itemize}
and thus, $\lvert P_{2\weight{a}+1}\rvert= \sum_{k=1}^{m}
\binom{2\weight{a}-\sum_{l=1}^k a_l +m-k}{m-k} $.

The reader can easily prove that if $z_i\in P_{2\weight{a}+1}$ then
for all $j>i$ with $\deg(z_j)=2\weight{a}+1$ it holds $z_j\in
P_{2\weight{a}+1}$. The details can be found in \cite{Tesi}. Thus,
$r^*(t)=\lvert\{z_i:\deg(z_i)\leq 2\weight{a}+1\}\rvert-\lvert
P_{2\weight{a}+1}\rvert.$

Now,
\begin{eqnarray*}
r^*(t)&=&\binom{2 \weight{a}+1+m}{m}- \lvert P_{2\weight{a}+1}\rvert
\\&=&
\binom{2 \weight{a}+1+m}{m}- \sum_{k=1}^m\binom{2 \weight{a}-
\sum_{l=1}^k a_l+m-k}{m-k}
\end{eqnarray*}

and

\begin{eqnarray*}
\tilde{r}^*(t)&=&r^*(t)-\lvert P_{2\weight{a}}\rvert\\
&=& r^*(t) -1-\sum_{k=1}^{m-1} \sum_{j=k}^{m-1}
\binom{2\weight{a}-2-\sum_{l=1}^j a_l -\sum_{l=1}^k a_l+m-j}{m-j}\\
& &-\lvert\{k : \weight{a}-\sum_{l=1}^k a_l>0\}\rvert.
\end{eqnarray*}
\end{enumerate}
\end{proof}

\begin{remark}
{\rm For $m\ll t$, $r(t)$ is $o(t^m)$, while $r^*(t)$ and
$\tilde{r}^*(t)$ are $o(t)$. Indeed, notice that
$\binom{a+b}{b}=\frac{a\cdot (a-1)\cdot\dots\cdot(a-b+1)}{b!}$ is
$o(a^b)$ if $a\gg b$. Now, for $m\ll t$, $r(t)$ is $o(t^m)$ while
$r^*(t)$ and $\tilde{r}^*(t)$ are $o((\deg(z_t))^m)$. On the other
hand, $\deg(z_t)$ is $o(t^{1/m})$, since all polynomials of degree
$k$ have order from $\binom{m+k-1}{m}$ to $\binom{m+k}{m}-1$.}
\end{remark}

\begin{figure}
\label{fig:red1}
\begin{center}
\setlength{\unitlength}{1cm}\font\graphicfont=cmr10 scaled 1000
\def\symbolr{\makebox(0,0){$\times$}}
\def\symboltr{\makebox(0,0){$+$}}
\def\symbolsr{\circle*{0.20}}
\def\symboltsr{\circle{0.20}}
\def\legendred{{\shortstack[c]{$\times$ $r(t)$\phantom{$r(t)$}
$+$ $\widetilde{r}(t)$\\ 
$\bullet$ $r^*(t)$\phantom{$r(t)$}
$\circ$ $\widetilde{r}^*(t)$
}}}

\fbox{\begin{picture}(11.500000,7.000000)
\put(0,6.700000){\parbox{11.500000cm}{\graphicfont{\begin{center}Reed-Muller structure with $m=3$\end{center}}}}
\put(1.080000,1.980000){\vector(1,0){9.845000}}
\put(10.925000,1.980000){\makebox(0.575000,0){$t$}}
\put(1.080000,1.980000){\line(0,1){4.420000}}
\put(1.080000,1.980000){\line(-1,0){0.080000}}
\put(0,1.780000){\makebox(1.000000,0.400000){$0$}}
\put(1.080000,4.422156){\line(-1,0){0.080000}}
\put(0,4.222156){\makebox(1.000000,0.400000){$25000$}}
\put(0,0){\makebox(11.500000,1.500000)\legendred}
\put(1.080000,1.980000){\line(0,-1){0.080000}}
\put(0.580000,1.500000){\makebox(1.000000,0.400000){$0$}}
\put(4.070323,1.980000){\line(0,-1){0.080000}}
\put(3.570323,1.500000){\makebox(1.000000,0.400000){$10$}}
\put(7.060645,1.980000){\line(0,-1){0.080000}}
\put(6.560645,1.500000){\makebox(1.000000,0.400000){$20$}}
\put(10.050968,1.980000){\line(0,-1){0.080000}}
\put(9.550968,1.500000){\makebox(1.000000,0.400000){$30$}}
\put(1.080000,1.980000){\symboltsr}
\put(1.080000,1.980000){\symbolsr}
\put(1.080000,1.980000){\symboltr}
\put(1.080000,1.980000){\symbolr}
\put(1.379032,1.980391){\symboltsr}
\put(1.379032,1.980391){\symbolsr}
\put(1.379032,1.980391){\symboltr}
\put(1.379032,1.980391){\symbolr}
\put(1.678065,1.980782){\symboltsr}
\put(1.678065,1.981075){\symbolsr}
\put(1.678065,1.981270){\symboltr}
\put(1.678065,1.981954){\symbolr}
\put(1.977097,1.981270){\symboltsr}
\put(1.977097,1.981368){\symbolsr}
\put(1.977097,1.982442){\symboltr}
\put(1.977097,1.985470){\symbolr}
\put(2.276129,1.981954){\symboltsr}
\put(2.276129,1.981954){\symbolsr}
\put(2.276129,1.983712){\symboltr}
\put(2.276129,1.991722){\symbolr}
\put(2.575161,1.982344){\symboltsr}
\put(2.575161,1.983517){\symbolsr}
\put(2.575161,1.985177){\symboltr}
\put(2.575161,2.001491){\symbolr}
\put(2.874194,1.982735){\symboltsr}
\put(2.874194,1.983614){\symbolsr}
\put(2.874194,1.987229){\symboltr}
\put(2.874194,2.015558){\symbolr}
\put(3.173226,1.983419){\symboltsr}
\put(3.173226,1.984005){\symbolsr}
\put(3.173226,1.988401){\symboltr}
\put(3.173226,2.034704){\symbolr}
\put(3.472258,1.983810){\symboltsr}
\put(3.472258,1.984103){\symbolsr}
\put(3.472258,1.990746){\symboltr}
\put(3.472258,2.059712){\symbolr}
\put(3.771290,1.984396){\symboltsr}
\put(3.771290,1.984494){\symbolsr}
\put(3.771290,1.992797){\symboltr}
\put(3.771290,2.091362){\symbolr}
\put(4.070323,1.985470){\symboltsr}
\put(4.070323,1.985470){\symbolsr}
\put(4.070323,1.994848){\symboltr}
\put(4.070323,2.130437){\symbolr}
\put(4.369355,1.985861){\symboltsr}
\put(4.369355,1.988303){\symbolsr}
\put(4.369355,1.996607){\symboltr}
\put(4.369355,2.177717){\symbolr}
\put(4.668387,1.986252){\symboltsr}
\put(4.668387,1.988401){\symbolsr}
\put(4.668387,1.999830){\symboltr}
\put(4.668387,2.233984){\symbolr}
\put(4.967420,1.986643){\symboltsr}
\put(4.967420,1.988499){\symbolsr}
\put(4.967420,2.001296){\symboltr}
\put(4.967420,2.299923){\symbolr}
\put(5.266452,1.987522){\symboltsr}
\put(5.266452,1.988987){\symbolsr}
\put(5.266452,2.004031){\symboltr}
\put(5.266452,2.376606){\symbolr}
\put(5.565484,1.987913){\symboltsr}
\put(5.565484,1.989085){\symbolsr}
\put(5.565484,2.006961){\symboltr}
\put(5.565484,2.464524){\symbolr}
\put(5.864516,1.988303){\symboltsr}
\put(5.864516,1.989183){\symbolsr}
\put(5.864516,2.009306){\symboltr}
\put(5.864516,2.564555){\symbolr}
\put(6.163549,1.989085){\symboltsr}
\put(6.163549,1.989671){\symbolsr}
\put(6.163549,2.011064){\symboltr}
\put(6.163549,2.677480){\symbolr}
\put(6.462581,1.989476){\symboltsr}
\put(6.462581,1.989769){\symbolsr}
\put(6.462581,2.015460){\symboltr}
\put(6.462581,2.804081){\symbolr}
\put(6.761613,1.990159){\symboltsr}
\put(6.761613,1.990257){\symbolsr}
\put(6.761613,2.016632){\symboltr}
\put(6.761613,2.945140){\symbolr}
\put(7.060645,1.991722){\symboltsr}
\put(7.060645,1.991722){\symbolsr}
\put(7.060645,2.020442){\symboltr}
\put(7.060645,3.101438){\symbolr}
\put(7.359678,1.992113){\symboltsr}
\put(7.359678,1.996216){\symbolsr}
\put(7.359678,2.023373){\symboltr}
\put(7.359678,3.273757){\symbolr}
\put(7.658710,1.992504){\symboltsr}
\put(7.658710,1.996314){\symbolsr}
\put(7.658710,2.025424){\symboltr}
\put(7.658710,3.462877){\symbolr}
\put(7.957742,1.992895){\symboltsr}
\put(7.957742,1.996411){\symbolsr}
\put(7.957742,2.028062){\symboltr}
\put(7.957742,3.669582){\symbolr}
\put(8.256775,1.993285){\symboltsr}
\put(8.256775,1.996509){\symbolsr}
\put(8.256775,2.032751){\symboltr}
\put(8.256775,3.894651){\symbolr}
\put(8.555807,1.994360){\symboltsr}
\put(8.555807,1.997095){\symbolsr}
\put(8.555807,2.035095){\symboltr}
\put(8.555807,4.138866){\symbolr}
\put(8.854839,1.994751){\symboltsr}
\put(8.854839,1.997193){\symbolsr}
\put(8.854839,2.037733){\symboltr}
\put(8.854839,4.403010){\symbolr}
\put(9.153871,1.995141){\symboltsr}
\put(9.153871,1.997290){\symbolsr}
\put(9.153871,2.040956){\symboltr}
\put(9.153871,4.687765){\symbolr}
\put(9.452904,1.995532){\symboltsr}
\put(9.452904,1.997388){\symbolsr}
\put(9.452904,2.044766){\symboltr}
\put(9.452904,4.994207){\symbolr}
\put(9.751936,1.996509){\symboltsr}
\put(9.751936,1.997974){\symbolsr}
\put(9.751936,2.046524){\symboltr}
\put(9.751936,5.322824){\symbolr}
\put(10.050968,1.996900){\symboltsr}
\put(10.050968,1.998072){\symbolsr}
\put(10.050968,2.052092){\symboltr}
\put(10.050968,5.674494){\symbolr}
\put(10.350000,1.997290){\symboltsr}
\put(10.350000,1.998170){\symbolsr}
\put(10.350000,2.053265){\symboltr}
\put(10.350000,6.050000){\symbolr}
\end{picture}}
\end{center}
\caption{Redundancy values of Reed-Muller standard codes and all improved codes.}
\end{figure}
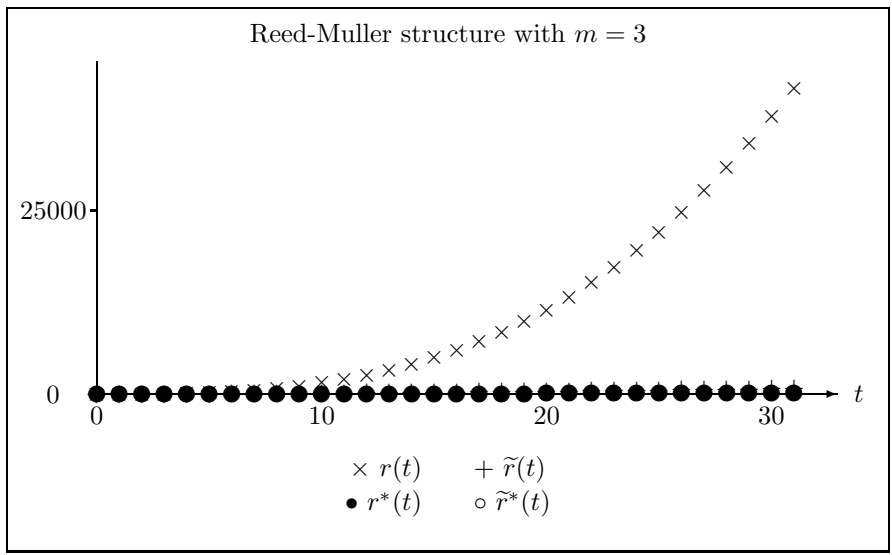

Let $m=3$. In Figure~\ref{fig:red1} we plot $r(t)$, $\tilde{r}(t)$,
${r}^*(t)$ and $\tilde{r}^*(t)$ as a function of $t$ for the first
values of $t$.
Notice that for all $t$, $r(t)$ is $o(t^3)$ while ${r}^*(t)$ and
$\tilde{r}^*(t)$ are $o(t)$. The function $\tilde{r}(t)$ seems to be
also $o(t)$.

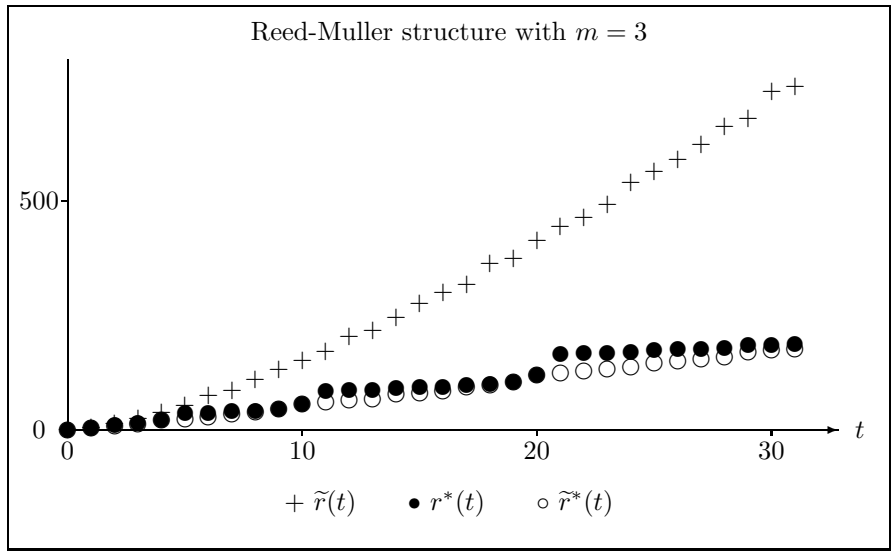
\begin{figure}
\label{fig:red2}
\begin{center}
\setlength{\unitlength}{1cm}\font\graphicfont=cmr10 scaled 1000
\def\symbolr{\makebox(0,0){$\times$}}
\def\symboltr{\makebox(0,0){$+$}}
\def\symbolsr{\circle*{0.20}}
\def\symboltsr{\circle{0.20}}
\def\legendred{{\shortstack[c]{$+$ $\widetilde{r}(t)$\phantom{$r(t)$}
$\bullet$ $r^*(t)$\phantom{$r(t)$}
$\circ$ $\widetilde{r}^*(t)$
}}}

\fbox{\begin{picture}(11.500000,7.000000)
\put(0,6.700000){\parbox{11.500000cm}{\graphicfont{\begin{center}Reed-Muller structure with $m=3$\end{center}}}}
\put(0.680000,1.480000){\vector(1,0){10.245000}}
\put(10.925000,1.480000){\makebox(0.575000,0){$t$}}
\put(0.680000,1.480000){\line(0,1){4.920000}}
\put(0.680000,1.480000){\line(-1,0){0.080000}}
\put(0,1.280000){\makebox(0.600000,0.400000){$0$}}
\put(0.680000,4.526667){\line(-1,0){0.080000}}
\put(0,4.326667){\makebox(0.600000,0.400000){$500$}}
\put(0,0){\makebox(11.500000,1.000000)\legendred}
\put(0.680000,1.480000){\line(0,-1){0.080000}}
\put(0.380000,1.000000){\makebox(0.600000,0.400000){$0$}}
\put(3.799355,1.480000){\line(0,-1){0.080000}}
\put(3.499355,1.000000){\makebox(0.600000,0.400000){$10$}}
\put(6.918710,1.480000){\line(0,-1){0.080000}}
\put(6.618710,1.000000){\makebox(0.600000,0.400000){$20$}}
\put(10.038065,1.480000){\line(0,-1){0.080000}}
\put(9.738065,1.000000){\makebox(0.600000,0.400000){$30$}}
\put(0.680000,1.480000){\symboltsr}
\put(0.680000,1.480000){\symbolsr}
\put(0.680000,1.480000){\symboltr}
\put(0.991935,1.504373){\symboltsr}
\put(0.991935,1.504373){\symbolsr}
\put(0.991935,1.504373){\symboltr}
\put(1.303871,1.528747){\symboltsr}
\put(1.303871,1.547027){\symbolsr}
\put(1.303871,1.559213){\symboltr}
\put(1.615806,1.559213){\symboltsr}
\put(1.615806,1.565307){\symbolsr}
\put(1.615806,1.632333){\symboltr}
\put(1.927742,1.601867){\symboltsr}
\put(1.927742,1.601867){\symbolsr}
\put(1.927742,1.711547){\symboltr}
\put(2.239677,1.626240){\symboltsr}
\put(2.239677,1.699360){\symbolsr}
\put(2.239677,1.802947){\symboltr}
\put(2.551613,1.650613){\symboltsr}
\put(2.551613,1.705453){\symbolsr}
\put(2.551613,1.930907){\symboltr}
\put(2.863548,1.693267){\symboltsr}
\put(2.863548,1.729827){\symbolsr}
\put(2.863548,2.004027){\symboltr}
\put(3.175484,1.717640){\symboltsr}
\put(3.175484,1.735920){\symbolsr}
\put(3.175484,2.150267){\symboltr}
\put(3.487419,1.754200){\symboltsr}
\put(3.487419,1.760293){\symbolsr}
\put(3.487419,2.278227){\symboltr}
\put(3.799355,1.821227){\symboltsr}
\put(3.799355,1.821227){\symbolsr}
\put(3.799355,2.406187){\symboltr}
\put(4.111290,1.845600){\symboltsr}
\put(4.111290,1.997933){\symbolsr}
\put(4.111290,2.515867){\symboltr}
\put(4.423226,1.869973){\symboltsr}
\put(4.423226,2.004027){\symbolsr}
\put(4.423226,2.716947){\symboltr}
\put(4.735161,1.894347){\symboltsr}
\put(4.735161,2.010120){\symbolsr}
\put(4.735161,2.808347){\symboltr}
\put(5.047097,1.949187){\symboltsr}
\put(5.047097,2.040587){\symbolsr}
\put(5.047097,2.978960){\symboltr}
\put(5.359032,1.973560){\symboltsr}
\put(5.359032,2.046680){\symbolsr}
\put(5.359032,3.161760){\symboltr}
\put(5.670968,1.997933){\symboltsr}
\put(5.670968,2.052773){\symbolsr}
\put(5.670968,3.308000){\symboltr}
\put(5.982903,2.046680){\symboltsr}
\put(5.982903,2.083240){\symbolsr}
\put(5.982903,3.417680){\symboltr}
\put(6.294839,2.071053){\symboltsr}
\put(6.294839,2.089333){\symbolsr}
\put(6.294839,3.691880){\symboltr}
\put(6.606774,2.113707){\symboltsr}
\put(6.606774,2.119800){\symbolsr}
\put(6.606774,3.765000){\symboltr}
\put(6.918710,2.211200){\symboltsr}
\put(6.918710,2.211200){\symbolsr}
\put(6.918710,4.002640){\symboltr}
\put(7.230645,2.235573){\symboltsr}
\put(7.230645,2.491493){\symbolsr}
\put(7.230645,4.185440){\symboltr}
\put(7.542581,2.259947){\symboltsr}
\put(7.542581,2.497587){\symbolsr}
\put(7.542581,4.313400){\symboltr}
\put(7.854516,2.284320){\symboltsr}
\put(7.854516,2.503680){\symbolsr}
\put(7.854516,4.477920){\symboltr}
\put(8.166452,2.308693){\symboltsr}
\put(8.166452,2.509773){\symbolsr}
\put(8.166452,4.770400){\symboltr}
\put(8.478387,2.375720){\symboltsr}
\put(8.478387,2.546333){\symbolsr}
\put(8.478387,4.916640){\symboltr}
\put(8.790323,2.400093){\symboltsr}
\put(8.790323,2.552427){\symbolsr}
\put(8.790323,5.081160){\symboltr}
\put(9.102258,2.424467){\symboltsr}
\put(9.102258,2.558520){\symbolsr}
\put(9.102258,5.282240){\symboltr}
\put(9.414194,2.448840){\symboltsr}
\put(9.414194,2.564613){\symbolsr}
\put(9.414194,5.519880){\symboltr}
\put(9.726129,2.509773){\symboltsr}
\put(9.726129,2.601173){\symbolsr}
\put(9.726129,5.629560){\symboltr}
\put(10.038065,2.534147){\symboltsr}
\put(10.038065,2.607267){\symbolsr}
\put(10.038065,5.976880){\symboltr}
\put(10.350000,2.558520){\symboltsr}
\put(10.350000,2.613360){\symbolsr}
\put(10.350000,6.050000){\symboltr}
\end{picture}}
\end{center}
\caption{Redundancy values of all improved codes.}
\end{figure}

Since $r(t)$ is much larger than the other three
functions, we cannot appreciate the differences between
$\tilde{r}(t)$, ${r}^*(t)$ and $\tilde{r}^*(t)$. If we only plot
$\tilde{r}(t)$, ${r}^*(t)$ and $\tilde{r}^*(t)$, (Figure~\ref{fig:red2})
then the relative
behavior of these functions becomes apparent.
In particular, $\tilde{r}^*(t)$ behaves as a smooth version of
$r^*(t)$.

\def\cprime{$'$}

\end{document}